\documentclass[floatfix,aps,preprint, superscriptaddress]{revtex4-2}
\usepackage[]{epsfig}
\usepackage{times,amsmath,amssymb}
\usepackage{graphicx,color}
\usepackage{hyperref}
\hypersetup{
colorlinks=true, 
urlcolor=blue
}

\begin{document}
\title{Automated Spin Readout Signal Analysis Using U-Net with Variable-Length Traces and Experimental Noise}

\author{Yui Muto}
\affiliation{Research Institute of Electrical Communication, Tohoku University, 2-1-1 Katahira, Aoba-ku, Sendai 980-8577, Japan}
\affiliation{Department of Electronic Engineering, Graduate School of Engineering, Tohoku University, Aoba 6-6-05, Aramaki, Aoba-Ku, Sendai 980-8579, Japan}

\author{Motoya Shinozaki}
\affiliation{WPI Advanced Institute for Materials Research, Tohoku University, 2-1-1 Katahira, Aoba-ku, Sendai 980-8577, Japan}
\affiliation{Research Center for Materials Nanoarchitechtonics (MANA), National Institute for Material Science (NIMS),
1-2-1 Sengen, Tsukuba 305-0047, Japan}

\author{Hideaki Yuta}
\affiliation{SANKEN, Osaka University, 8-1 Mihogaoka Ibaraki Osaka 567-0047, Japan}

\author{Tatsuo Tsuzuki}
\affiliation{SANKEN, Osaka University, 8-1 Mihogaoka Ibaraki Osaka 567-0047, Japan}

\author{Kotaro Taga}
\affiliation{SANKEN, Osaka University, 8-1 Mihogaoka Ibaraki Osaka 567-0047, Japan}

\author{Akira Oiwa}
\affiliation{SANKEN, Osaka University, 8-1 Mihogaoka Ibaraki Osaka 567-0047, Japan}

\author{Takafumi Fujita}
\affiliation{SANKEN, Osaka University, 8-1 Mihogaoka Ibaraki Osaka 567-0047, Japan}

\author{Tomohiro Otsuka}
\email[]{tomohiro.otsuka@tohoku.ac.jp}
\affiliation{WPI Advanced Institute for Materials Research, Tohoku University, 2-1-1 Katahira, Aoba-ku, Sendai 980-8577, Japan}
\affiliation{Research Institute of Electrical Communication, Tohoku University, 2-1-1 Katahira, Aoba-ku, Sendai 980-8577, Japan}
\affiliation{Department of Electronic Engineering, Graduate School of Engineering, Tohoku University, Aoba 6-6-05, Aramaki, Aoba-Ku, Sendai 980-8579, Japan}
\affiliation{Center for Science and Innovation in Spintronics, Tohoku University, 2-1-1 Katahira, Aoba-ku, Sendai 980-8577, Japan}
\affiliation{Research Center for Materials Nanoarchitechtonics (MANA), National Institute for Material Science (NIMS),
1-2-1 Sengen, Tsukuba 305-0047, Japan}
\affiliation{Center for Emergent Matter Science, RIKEN, 2-1 Hirosawa, Wako, Saitama 351-0198, Japan}

\date{\today}

\begin{abstract}
Single-shot spin-state discrimination is essential for semiconductor spin qubits, but conventional threshold-based analysis of spin readout traces becomes unreliable under noisy conditions.
Although recent neural-network-based methods improve robustness against experimental noise, they are sensitive to training conditions, restricted to fixed-length inputs, and limited to trace-level outputs without explicit temporal localization of transition events.
In this work, we apply a U-Net architecture to spin readout signal analysis by formulating transition-event detection as a point-wise segmentation task in one-dimensional time-series data.
The fully convolutional structure enables direct processing of variable-length traces.
Point-wise and sample-wise evaluations demonstrate low readout error rates and high classification accuracy without retraining.
The proposed method generalizes well to previously-unseen trace lengths and experimental non-Gaussian noise, outperforming a conventional threshold-based approach and providing a robust and practical solution for automated spin readout signal analysis.
\end{abstract}

\maketitle

\section{Introduction}\label{sec:intro}

To utilize semiconductor spin qubits in quantum computers, it is essential to read out the spin state with high accuracy in a single-shot measurement\cite{Elzerman2004}. 
In this scheme, the spin state is mapped onto a charge state through spin-to-charge conversion~\cite{Ono2002}, where the presence or absence of a charge transition event in the charge sensor signal~\cite{Field1993} reflects the underlying spin state.
However, noise in the charge sensor signals limits the performance of spin-state estimation using conventional threshold-based methods~\cite{Shinozaki2025}.

To address these challenges, previous studies have proposed approaches based on both hardware and software improvements. 
The former includes low-noise amplifiers~\cite{Yamamoto2008,Cochrane2022, Kass2023} and feedback systems~\cite{Nakajima2021, Fujiwara2023, Wen2025}, while the latter refers to Bayesian~\cite{Shinozaki2025} and likelihood-based approaches~\cite{Mizokuchi2020, Mizokuchi_2026}.
As a software-based approach, machine learning (ML) is also a promising tool. 
In the field of semiconductor quantum dots, ML-based techniques have been mainly employed to address scalability issues associated with device tuning and operation~\cite{Kalantre2019, 
Zwolak2023_Colloquium, Schuff2024, muto2024, Muto2025, Tsuzuki2026}.
Their application has been extended to spin readout, where convolutional neural networks (CNNs) have been used to automatically classify noisy spin readout signals~\cite{Y_Matsumoto_npjQI2021, Struck2021RobustFastPostProcessing, Oakes2023}.
For example, it has been demonstrated that a model trained on output traces from radio-frequency reflectometry measurements~\cite{reilly2007fast} can automatically determine whether a given trace contains a charge transition event~\cite{Y_Matsumoto_npjQI2021}.

On the other hand, three challenges can be considered when applying this approach to experimental environments.
First, there is a dependence on experimental conditions, as the model is trained using measured traces and may require retraining or fine-tuning to adapt to changes in experimental parameters.
Second, the model is restricted to fixed-length inputs, making it difficult to apply to traces of different lengths without redesigning the model architecture.
Third, the black-box nature of the model presents an additional issue. The model output indicates only the presence or absence of a transition event, without explicitly specifying which temporal regions of the trace are identified as transition events, making detailed verification of the results difficult.

In light of these challenges, there is a need for flexible spin signal analysis methods that remain robust under varying experimental conditions. 
In this context, studies applying U-Net to time-series anomaly detection \cite{Wen2019TimeSeriesAnomalyDetection, Perslev2019UTime} have shown that U-Net can effectively extract characteristic patterns from time-series signals, providing useful insights for the present work, despite being conducted in a different application domain.
U-Net has a fully convolutional architecture and integrates features at different scales (multi-scale features) through skip connections between corresponding layers of the encoder and decoder. This structure enables highly accurate localization by simultaneously exploiting both local and global features \cite{ronneberger2015u}. 
Furthermore, the output length of U-Net matches the input length, and the model is designed to output, for each point in the time series, the predicted probability of belonging to the anomalous class~\cite{Long2015FCN}. 
Compared with CNN-based approaches that classify an entire trace into a single class, this point-wise prediction provides more detailed information and is also useful for validating the plausibility of the analysis results.

In this study, we apply U-Net to spin readout signal analysis to address three key limitations including condition dependence, fixed-length input constraints, and the black-box nature of the models. 
By mapping the anomalous class in time-series anomaly detection to locally occurring transition events in spin readout traces, we position U-Net as a transition-event detector. 
In addition, its fully convolutional architecture enables a variable-length model that can be applied to traces of different lengths without retraining.

\begin{figure*}[ht]
\centering
\includegraphics[width=\textwidth]{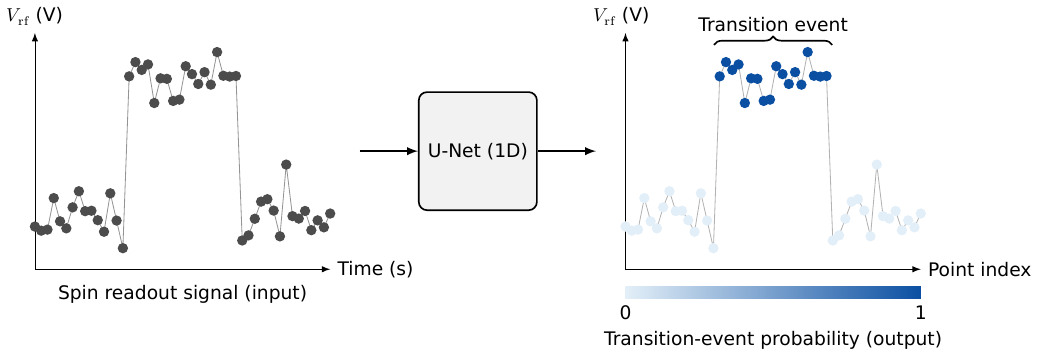}
\caption{Schematic overview of transition-event detection using U-Net employed in this study. Here, $V_\mathrm{rf}$ is the radio-reflectometry signal. A one-dimensional spin readout trace is input into U-Net, which outputs, for each sample point, the probability of belonging to a transition event. The proposed U-Net model supports variable-length inputs.}
\label{fig5_overflow}
\end{figure*}

A schematic overview of the proposed method is shown in Fig.~\ref{fig5_overflow}. 
In this paper, a single trace is defined as one sample, while each point along the time series constituting a trace is defined as a sample point. 
A one-dimensional spin readout trace is input into U-Net, which is designed and trained using simulated data to output, for each sample point, the probability of belonging to a transition event.
This point-wise probability output enables direct visualization of the temporal regions identified by the model as transition events on the trace, thereby alleviating the black-box nature of the model and facilitating validation of the analysis results. Subsequently, the discrimination performance of the trained U-Net model is systematically evaluated from both point-wise and sample-wise perspectives.

\section{Dataset Construction for Training and Evaluation of U-Net}
This section describes the construction of the datasets used for training and evaluating U-Net for spin readout signal analysis. In this study, either simulated noise or experimental noise is used as the noise component, while all pulse signals corresponding to transition events are generated by simulation.
In experimental measurements, electron dynamics are inherently stochastic, making it difficult to precisely identify the true locations of transition events and thus to assign accurate labels. By contrast, generating transition-event pulses through simulation enables the construction of large-scale training and evaluation datasets with precisely labeled transition-event locations \cite{Struck2021RobustFastPostProcessing}.

\begin{figure*}[ht]
\centering
\includegraphics[width=\textwidth]{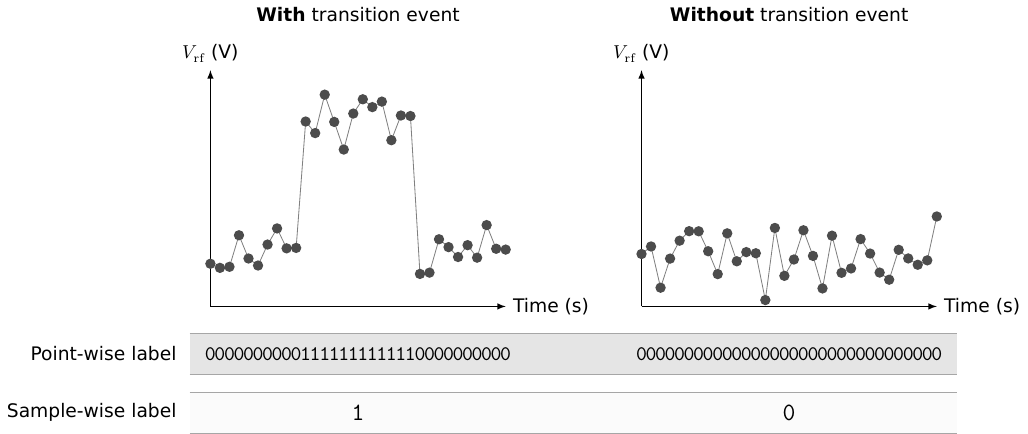}
\caption{Illustration of point-wise and sample-wise labeling used in this study.
(top) Traces with and without a transition event;
(middle) Point-wise labels;
(bottom) Sample-wise labels.}
\label{fig5_dual_labeling}
\end{figure*}

Model performance is evaluated from both point-wise and sample-wise perspectives.
To this end, two types of labels are defined for each trace, as schematically illustrated in Fig.~\ref{fig5_dual_labeling}.
Point-wise labels indicate the temporal locations of transition events at each sample point, whereas sample-wise labels represent the presence or absence of at least one transition event within a trace.
This labeling strategy enables a consistent evaluation of transition-event detection accuracy at the point level and spin-state discrimination performance at the trace level.

Point-wise labels are used for both training and evaluation of the U-Net model.
By contrast, sample-wise labels are not used for model training but are introduced solely for post hoc evaluation, providing a performance measure relevant to final spin-state discrimination in practical operation.

\subsection{Training Data Generation}\label{sec:training_data}
First, Gaussian noise with a mean of zero and a standard deviation randomly sampled from a uniform distribution in the range of 0.1 to 3 is generated, and a total of 96{,}000 noise traces are used as data without transition events.

Data with transition events are generated by superimposing a transition pulse of height 1 onto the same noise traces. 
The transition pulses are generated using a state-transition model based on a Markov chain, with three different numbers of tunneling attempts per sweep time: 0.4, 4, and 40.
These conditions correspond to slow, intermediate, and fast tunneling rates~$\Gamma$,
$2.0\times10^{4}\,\mathrm{s}^{-1}$, $2.0\times10^{5}\,\mathrm{s}^{-1}$, and $2.0\times10^{6}\,\mathrm{s}^{-1}$,
respectively, and are introduced to improve the generality of the model under different experimental conditions.
The sweep time is fixed at $T_{\mathrm{sweep}} = 20~\mu\mathrm{s}$,
while the data length $L$ is varied as described below.
Accordingly, the sampling interval $\Delta t = T_{\mathrm{sweep}}/L$
differs depending on the data length, and the discrete-time transition
probability is determined from the tunneling rate~$\Gamma$ using this $\Delta t$.
Because tunneling is stochastic, no transition may occur within a sweep time; however, during transition-pulse generation, only pulses containing at least one transition-event point are retained.
In this study, we consider Elzerman-type single-shot spin readout, in which a spin-dependent tunneling event occurs. 
This event appears as a single pulse in the time-domain trace.
Following this procedure, 96{,}000 traces with transition events are generated.

The data length $L$ is evenly set to six values: 64, 128, 256, 512, 1024, and 2048.
For training data generation, point-wise labels are encoded as binary vectors (1 for transition-event points and 0 otherwise), consistent with the example shown in Fig.~\ref{fig5_dual_labeling}.
All generated data are split into training, validation, and test sets in a 7:2:1 ratio.
The training set is used to update the model weights, the validation set is used to monitor overfitting during training, and the test set is used for evaluation.
All data are standardized using the mean $\mu_{\mathrm{train}}$ and standard deviation $\sigma_{\mathrm{train}}$ calculated from the training set (z-score normalization).
In this study, the U-Net model is trained only once using the training data described in this subsection, and all subsequent predictions and evaluations are performed solely by inference using this trained model, without any transfer learning or retraining.

\subsection{Evaluation Data Construction}\label{sec:eval_data}

\begin{figure*}[ht]
\centering
\includegraphics[width=\textwidth]{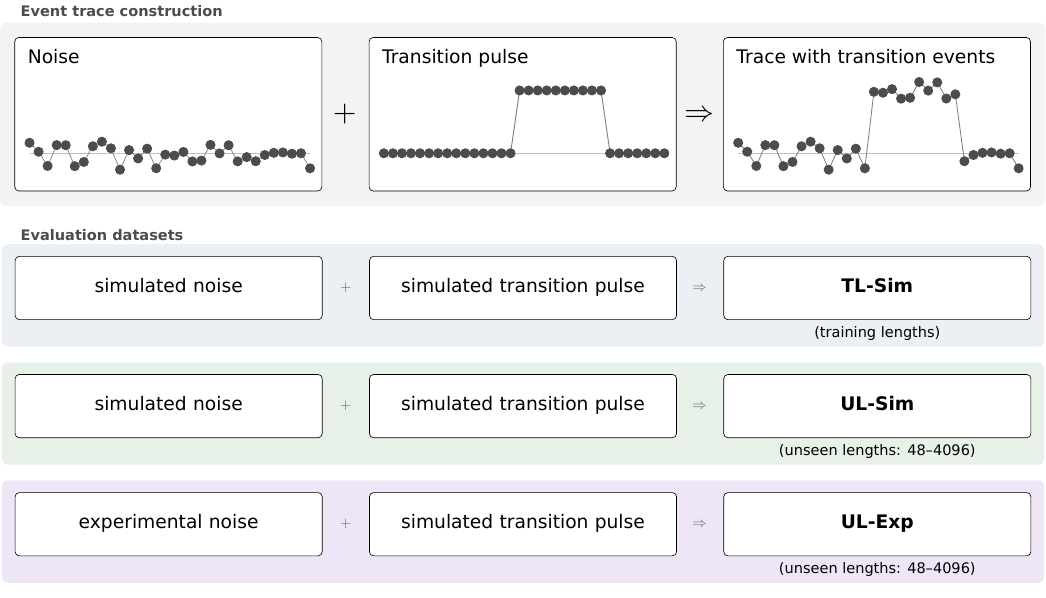}
\caption{
Overview of the three evaluation datasets (TL-Sim, UL-Sim, and UL-Exp).
The top row shows a common schematic of how traces with transition events are constructed by superimposing simulated transition pulses onto noise traces.
The panels below detail the dataset-specific composition for TL-Sim, UL-Sim, and UL-Exp, indicating whether the noise traces are simulated or experimentally acquired (the transition pulses are simulated in all cases).
TL and UL denote data lengths seen and unseen during the training phase, respectively.
The colored backgrounds are used consistently throughout this paper to identify each dataset and to link this schematic definition to the corresponding evaluation results.
}
\label{fig5_dataset_overview}
\end{figure*}

To systematically evaluate the generalization and robustness of the trained U-Net model, three types of evaluation datasets are prepared: Train-Length Simulation Data (TL-Sim), Unseen-Length Simulation Data (UL-Sim), and Unseen-Length Experimental-Noise Data (UL-Exp).
Using these datasets, the trained U-Net model is evaluated on TL-Sim to assess generalization to unseen samples generated under training-matched conditions, while robustness to variable-length inputs and to experimental noise is examined using UL-Sim and UL-Exp, respectively (see Fig.~\ref{fig5_dataset_overview}).

For both UL-Sim and UL-Exp, samples with and without transition events are generated as paired data by changing only the presence or absence of transition pulses added to the same noise trace.
As a result, the numbers of samples in the two classes are balanced with a 1:1 ratio.
For TL-Sim, only samples containing transition events are used in the subsequent evaluations.

\subsubsection{Train-Length Simulation Data (TL-Sim)}

TL-Sim corresponds to the test dataset described in Sec.~\ref{sec:training_data}, which is generated using the same data-generation process and data lengths as the training data but is not used during training.
This dataset is used to evaluate the performance of the trained model on unseen samples generated under the same conditions as the training data.

\subsubsection{Unseen-Length Simulation Data (UL-Sim)}

UL-Sim is constructed to evaluate the robustness of the U-Net model to variable-length inputs.
For this dataset, simulated noise with a mean of zero and a standard deviation in the range of 0.1 to 1 is used, while the transition-pulse height (1) and tunneling rates
($2.0\times10^{4}\,\mathrm{s}^{-1}$, $2.0\times10^{5}\,\mathrm{s}^{-1}$, and $2.0\times10^{6}\,\mathrm{s}^{-1}$)
are set to be identical to those used for the training data.

A total of 440{,}000 traces are generated, with data lengths varied over a wide range from 48 to 4096.
This range includes intermediate values between the training data lengths (e.g., 192 and 384), values slightly offset from the training data lengths (e.g., 230 and 282), and significantly shorter or longer traces outside the training range (e.g., 48, 3072, and 4096).
All data are standardized using the same $\mu_{\mathrm{train}}$ and $\sigma_{\mathrm{train}}$ as those used for the training data.

\subsubsection{Unseen-Length Experimental-Noise Data (UL-Exp)}\label{sec:UL-Exp_preparation}
UL-Exp is constructed using experimental noise traces acquired by radio-frequency reflectometry in GaAs quantum dots~\cite{Tsuzuki2026}, onto which simulated transition pulses are superimposed.
The data lengths are varied in the same manner as in UL-Sim, covering a wide range from 48 to 4096, and a total of 1{,}446{,}400 traces are prepared.

This dataset is designed to simultaneously evaluate generalization with respect to data length and robustness against experimental noise.
For each sample, the transition amplitude is determined from the ratio of the standard deviation of the experimental noise to a prescribed noise-level value in the range of 0.1 to 1, and a calibration is performed to match the noise standard deviation and transition amplitude on a per-sample basis.
All data are standardized using the same $\mu_{\mathrm{train}}$ and $\sigma_{\mathrm{train}}$ as those used for the training data.

\noindent\textbf{Preparation Procedure}\par\vspace{0.3em}
In this paragraph, the preparation procedure for UL-Exp is quantitatively described using mathematical formulations.
UL-Exp is an evaluation-only dataset constructed by extracting experimental noise traces with data lengths not used for training and superimposing simulated transition pulses onto them.
This dataset is designed to simultaneously evaluate the generalization performance with respect to data length, as assessed using UL-Sim, and the robustness against experimental noise.

First, an amplitude signal is computed from the In- and Quadrature-phase signals obtained by reflectometry measurements and is regarded as an experimental noise trace $\mathbf{n}(t)$.
After removing statistically identical duplicate traces, a set of unique experimental noise traces is constructed and randomly shuffled.

Next, for each data length $L$ ranging from $48$ to $4096$,
the entire experimental noise traces $\mathbf{n}(t)$ are independently segmented
into short sub-traces $\mathbf{n}_i \in \mathbb{R}^{L}$,
where $i$ denotes the sample index.
As a result, the number of extracted sub-traces differs depending on the data length.
To enable fair performance comparisons under identical conditions across different data lengths,
random sampling is then applied so that the number of samples is matched
to the minimum sample count among all data lengths.

For each sub-trace, the same three numbers of tunneling rates ($2.0\times10^{4}\,\mathrm{s}^{-1}$, $2.0\times10^{5}\,\mathrm{s}^{-1}$, and $2.0\times10^{6}\,\mathrm{s}^{-1}$) as used previously are assigned in an approximately balanced manner so that the number of samples does not become biased.
Let $\sigma_{\mathrm{noise}, i}$ denote the standard deviation of the experimental noise sub-trace $\mathbf{n}_i$ in the raw scale.
A noise level $\mathrm{NL}_i$ is then sampled from a uniform distribution in the range of 0.1 to 1.
At this point, the transition-pulse amplitude $A_i$ for each sample is determined as
\begin{equation}
  A_i = \frac{\sigma_{\mathrm{noise}, i}}{\mathrm{NL}_i}.
\end{equation}
Here, $\mathrm{NL}_i$ is a dimensionless quantity representing the ratio of the noise intensity to the transition amplitude.

The transition pulse $\mathbf{p}_i \in \mathbb{R}^{L}$ is generated by simulation using a state-transition model based on a Markov chain, according to the assigned tunneling rate and the data length $L$.
Here, $\mathbf{p}_i$ is defined as a normalized pulse with unit height at transition-event locations and zero elsewhere, and is constrained to contain at least one transition-event point for all samples.
The resulting transition pulse is superimposed onto the experimental noise to generate a trace with transition events as
\begin{equation}
  \mathbf{d}^{\mathrm{raw}}_i = \mathbf{n}_i + A_i \mathbf{p}_i .
\end{equation}
At the same time, the noise-only trace
\begin{equation}
  \mathbf{u}^{\mathrm{raw}}_i = \mathbf{n}_i ,
\end{equation}
is retained.
Here, $\mathbf{d}^{\mathrm{raw}}_i$ corresponds to data with transition events, whereas $\mathbf{u}^{\mathrm{raw}}_i$ corresponds to data without transition events.

Next, baseline- and gain-based calibration is performed for each sample by applying
\begin{equation}
  \tilde{\mathbf{x}}_i
    = \left( \mathbf{x}^{\mathrm{raw}}_i - \mathrm{baseline}_i \right)\cdot \mathrm{gain}_i .
\end{equation}
Here, $\mathbf{x}^{\mathrm{raw}}_i$ denotes either $\mathbf{d}^{\mathrm{raw}}_i$ or $\mathbf{u}^{\mathrm{raw}}_i$, and $\mathrm{baseline}_i$ is defined as the mean value of the corresponding experimental noise sub-trace $\mathbf{n}_i$, which is used to correct offset components originating from the measurement circuitry.
In addition, $\mathrm{gain}_i$ is a scaling factor based on the transition amplitude and is defined as
\begin{equation}
  \mathrm{gain}_i = A_i^{-1} .
\end{equation}
Through this gain scaling, sample-dependent transition amplitudes $A_i$ are normalized, such that the normalized height of transition events is set to unity for all samples.
Note that this calibration procedure is not limited to synthetic traces constructed by superimposing simulated transition pulses onto experimentally acquired noise traces such as UL-Exp, but is also applicable to fully experimental traces containing transition events.
Even in this case, the baseline (noise mean) and the transition amplitude can be experimentally estimated from the data, enabling baseline correction and gain scaling based on these estimates.

Furthermore, standardization is applied using the mean $\mu_{\mathrm{train}}$ and standard deviation $\sigma_{\mathrm{train}}$ precomputed from the training data as
\begin{equation}
  \mathbf{x}^{\mathrm{scaled}}_i
    = \frac{\tilde{\mathbf{x}}_i - \mu_{\mathrm{train}}}{\sigma_{\mathrm{train}}} .
\end{equation}

We further examine the amplitude distribution to characterize the statistical properties of the experimental noise used in UL-Exp.
The distribution does not exhibit a Gaussian shape symmetric about the mean; instead, it shows a clear asymmetry, with a higher probability of values occurring below the mean.
A plausible explanation for this asymmetry is related to the operating point of the charge-sensing quantum dot.
Although the exact origin cannot be fully identified, the sensor dot may be biased away from the optimal point on its conductance peak.
A shift toward either the peak maximum or the tail of the conductance peak can lead to reduced sensitivity on one side of the signal response.
This behavior suggests that fluctuations in the local electrostatic environment of the sensor dot play a dominant role, indicating a non-ideal but realistic operating condition.
Such deviations from ideal Gaussian noise are common in practical quantum-dot measurements.
Therefore, the asymmetric noise distribution observed in UL-Exp reflects inherent experimental non-idealities and enables a more realistic evaluation of model robustness than evaluations based solely on simulated noise.

Through these procedures, UL-Exp is constructed as an evaluation dataset that simultaneously probes generalization with respect to data length and robustness against experimentally derived noise.

\section{U-Net Model Architecture and Training}
The architecture of the U-Net model used in this study is based on that of a prior study on time-series anomaly detection introduced in Sec.~\ref{sec:intro}~\cite{Wen2019TimeSeriesAnomalyDetection}.
An overview of the U-Net architecture used in this study is summarized in Table~\ref{tab:unet_architecture}.
Specifically, a one-dimensional U-Net with a four-level encoder--decoder structure and skip connections is employed.
At each resolution level, a convolution block (conv\_block) consisting of a 1D convolution with kernel size 3, followed by batch normalization and a ReLU activation, is applied.
In the encoder, two consecutive conv\_blocks are used at each level, followed by temporal downsampling using MaxPooling1D with a pooling factor of 4.
With this design, the encoder comprises four stages of downsampling with an overall downsampling factor of $4^{4}=256$, while the number of feature channels is progressively increased from 16 to 256.
In the decoder, the feature maps are symmetrically upsampled using UpSampling1D with the same factor, concatenated with the corresponding encoder features via skip connections, and refined by a 1D convolution followed by two conv\_blocks.
The final output layer is a $1\times1$ convolution followed by a softmax activation, producing point-wise probabilities for two classes at each time sample.

\begin{table}[t]
\centering
\caption{Architecture of the 1D U-Net model used in this study.}
\label{tab:unet_architecture}
\setlength{\tabcolsep}{4pt}
\begin{tabular}{lccc}
\hline
Stage & Operation & Kernel / Pool size & Channels \\
\hline
Input
& 1D signal
& --
& 1 \\

Enc-1
& conv\_block $\times2$ + MaxPool
& $3,\ 4$
& 16 \\

Enc-2
& conv\_block $\times2$ + MaxPool
& $3,\ 4$
& 32 \\

Enc-3
& conv\_block $\times2$ + MaxPool
& $3,\ 4$
& 64 \\

Enc-4
& conv\_block $\times2$ + MaxPool
& $3,\ 4$
& 128 \\

Bottleneck
& conv\_block $\times2$
& $3$
& 256 \\

Dec-4
& UpSample + Concat + conv\_block $\times2$
& $4,\ 3$
& 128 \\

Dec-3
& UpSample + Concat + conv\_block $\times2$
& $4,\ 3$
& 64 \\

Dec-2
& UpSample + Concat + conv\_block $\times2$
& $4,\ 3$
& 32 \\

Dec-1
& UpSample + Concat + conv\_block $\times2$
& $4,\ 3$
& 16 \\

Output
& $1\times1$ Conv + Softmax
& $1$
& 2 \\
\hline
\end{tabular}
\end{table}

In addition, to enable the model to handle variable-length traces, explicit length-alignment processing based on padding and cropping is introduced.
Since pooling and upsampling are performed over four levels, shape mismatches arise when the input length is not a multiple of the overall downsampling factor.
To avoid this issue, zero padding is applied to the right end of each input trace so that its length becomes a multiple of 256.
As a result, shape consistency across all layers is preserved throughout the encoder--decoder network.
After decoding, the output is cropped at the right end to match the original input length, enabling a one-to-one correspondence between each output sample and the corresponding input sample.
This length-alignment procedure allows the proposed U-Net to be applied in a unified and stable manner to time-series signals of varying lengths.
During training, variable-length traces are zero-padded to the maximum length within each batch so that all samples share the same length.

As the loss function, the soft Dice loss is adopted to quantify the overlap between the predicted transition-event region and the ground-truth region in the segmentation task.
A smaller soft Dice loss indicates better agreement between the prediction and the ground truth, and the model parameters are optimized by minimizing this loss during training.
The soft Dice loss is defined as
\begin{equation}
\text{soft Dice loss} = 1 -
\frac{2 \sum y_{\mathrm{true}} \, y_{\mathrm{pred}}}
{\sum \left( y_{\mathrm{true}}^2 + y_{\mathrm{pred}}^2 \right)} .
\end{equation}
Here, the loss is computed only for the positive class corresponding to transition events.
The variable $y_{\mathrm{true}} \in \{0,1\}$ denotes the ground-truth point-wise label, while
$y_{\mathrm{pred}} \in [0,1]$ represents the model-predicted probability of a transition event at each sample point.

To prevent padded regions from influencing the loss computation, a masking strategy is applied when calculating the soft Dice loss.
Specifically, padded regions are identified as points where the ground-truth labels correspond to neither class (i.e., both entries of the one-hot representation are zero), and these regions are excluded from the loss calculation.
This masking ensures that the soft Dice loss is evaluated only over valid portions of each trace, enabling consistent training for input data with varying sequence lengths.

\section{Evaluation of Data-Length Dependence and Point-Wise Classification Performance}
In this section, the generalization performance of the U-Net model with respect to data length is evaluated from the perspective of point-wise classification performance.
The evaluation is conducted using TL-Sim, which has the same data lengths as the training data; UL-Sim, which has data lengths different from those used for training; and UL-Exp, which has data lengths different from those used for training and incorporates experimental noise as the noise component.
This evaluation framework enables a unified assessment of the robustness to variations in data length, as well as the transfer performance from a simulation-noise environment to an experimental-noise environment.

The U-Net outputs, via a softmax function, a two-class probability distribution at each sample point, corresponding to the presence and absence of a transition event.
Point-wise binarization is performed by assigning each sample point to the class with the higher predicted probability. 
As an evaluation metric, the point-wise error rate (ER$_{\mathrm{point}}$), defined as follows, is used:
\begin{equation}
\mathrm{ER}_{\mathrm{point}}
= 1 - \frac{TP_{\mathrm{point}} + TN_{\mathrm{point}}}
{TP_{\mathrm{point}} + FP_{\mathrm{point}} + FN_{\mathrm{point}} + TN_{\mathrm{point}}} .
\end{equation}
Here, $TP_{\mathrm{point}}$ denotes the number of sample points that are correctly classified as having transition events,
and $TN_{\mathrm{point}}$ denotes the number of sample points that are correctly classified as having no transition events.
The quantity $FP_{\mathrm{point}}$ represents the number of sample points that are incorrectly classified as having transition events despite the absence of true transition events,
while $FN_{\mathrm{point}}$ represents the number of sample points that are incorrectly classified as having no transition events despite the presence of true transition events.
Note that $\mathrm{ER}_{\mathrm{point}}$ corresponds to $1$ minus the point-wise classification accuracy and thus directly reflects the fraction of misclassified sample points.

\begin{figure}
\begin{center}
  \includegraphics[width=\textwidth]{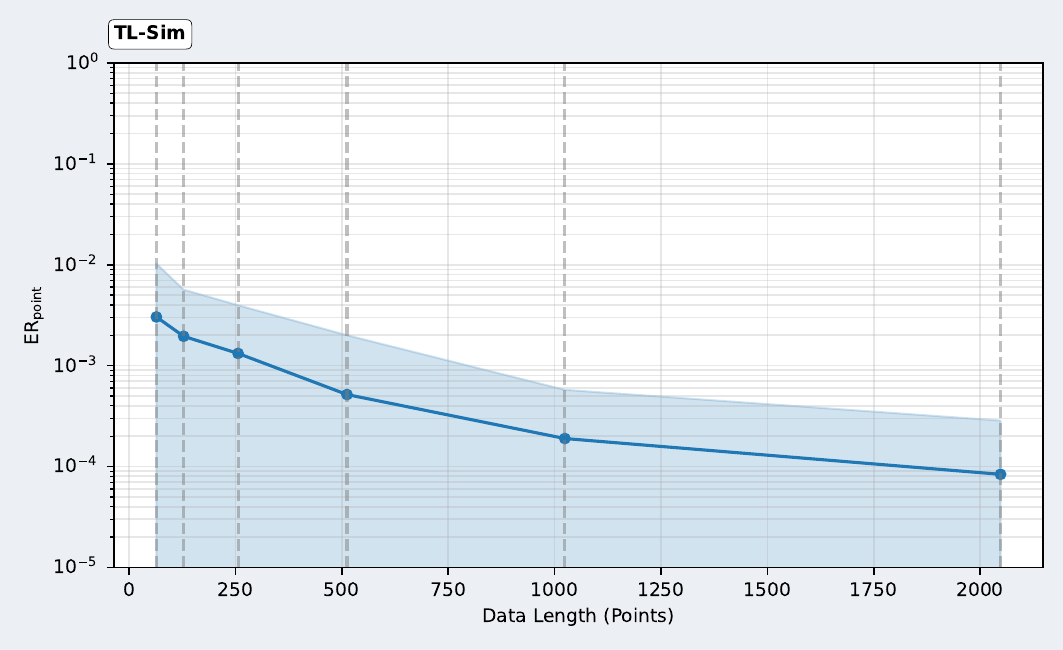}
  \caption{Point-wise error rate of U-Net evaluated on TL-Sim, which has the same data lengths as the training data.
Only samples with transition events from the TL-Sim dataset are included, and the noise level is restricted to
$0.2 \leq \text{noise level} < 0.3$.
For each data length, the number of samples is approximately 70; markers indicate the mean values and shaded bands represent the standard deviations.
The gray dashed lines denote the data lengths used during training.}
  \label{fig5_TL_Sim_Point_Wise_ER}
\end{center}
\end{figure}

First, the evaluation results obtained for TL-Sim, which has the same data lengths as the training data, are shown in Fig.~\ref{fig5_TL_Sim_Point_Wise_ER}.
The analysis uses only samples with transition events from the entire TL-Sim dataset, and the noise level is restricted to the range
$0.2 \leq \text{noise level} < 0.3$,
which corresponds to typical experimental conditions~\cite{Elzerman2004}.
For all data lengths, $\mathrm{ER}_{\mathrm{point}}$ remains below $10^{-2}$, indicating that high point-wise classification accuracy is maintained.
This result demonstrates that the limitation of fixed-length inputs, which was a major drawback of conventional methods, is effectively resolved.
Furthermore, while conventional CNN-based approaches output only the presence or absence of a transition event on a per-sample (per-trace) basis, the proposed method directly outputs transition-event predictions at the sample-point level.
As a result, point-wise error-rate evaluation becomes possible, and transition events can be identified with high accuracy, including their temporal locations, indicating that the black-box nature inherent in conventional approaches is successfully mitigated.

\begin{figure}
\begin{center}
  \includegraphics[width=\textwidth]{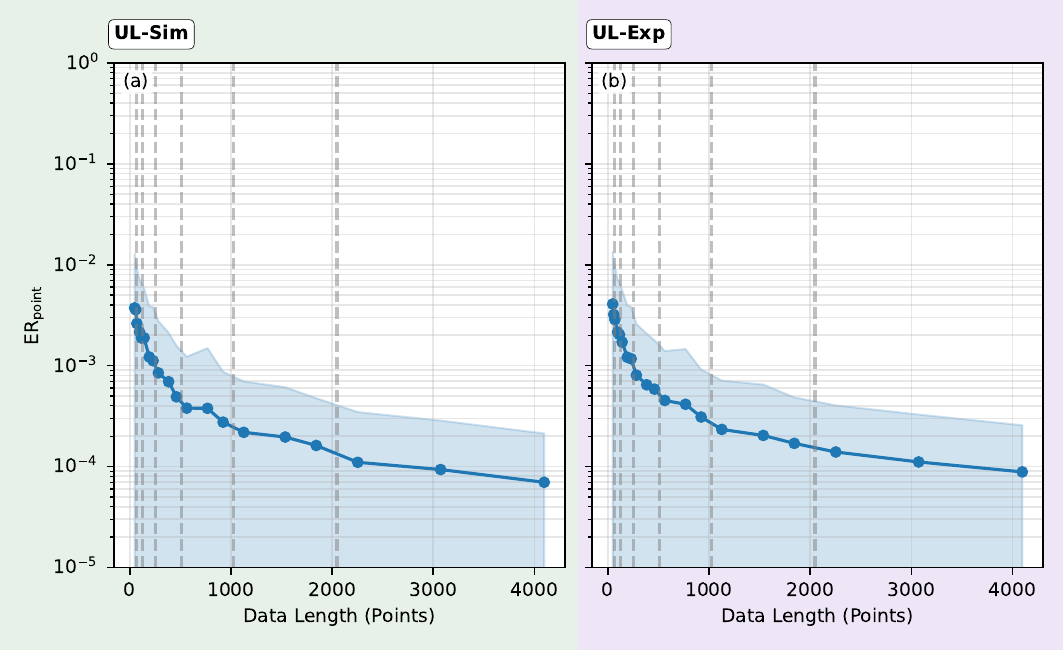}
  \caption{Point-wise error rate of the U-Net evaluated on untrained data lengths.
(a) UL-Sim, which consists of simulated noise traces with data lengths different from those used during training.
(b) UL-Exp, which incorporates experimentally acquired noise traces and data lengths unseen during training.
In both panels, only samples containing transition events are included, and the noise level is restricted to
$0.2 \leq \text{noise level} < 0.3$.
Markers indicate mean values and shaded bands represent standard deviations.
The numbers of samples are approximately 1200 for UL-Sim and 4000 for UL-Exp at each data length.
The gray dashed lines denote the data lengths used during training.
}\label{compare_overall_pointwise_ER_vs_length}
\end{center}
\end{figure}

Next, the results of the same evaluation performed on UL-Sim, which consists of data lengths different from those used for training, are shown in Fig.~\ref{compare_overall_pointwise_ER_vs_length}(a).
For all data lengths, $\mathrm{ER}_{\mathrm{point}}$ remains comparable to that obtained for TL-Sim, indicating that little to no performance degradation is observed even for data lengths unseen during training.
UL-Sim includes intermediate data lengths between those used for training, slightly offset lengths, as well as significantly shorter and longer traces outside the training range; nevertheless, U-Net exhibits stable classification performance across all these cases.
This robustness can be attributed to the fact that U-Net learns local features through convolutional filters and adopts a fully convolutional architecture that does not depend on the input length.
From these results, it is confirmed that U-Net possesses high generality with respect to unseen data lengths and effectively overcomes the condition dependence that has been a limitation of conventional approaches.

Furthermore, the results of the same evaluation performed on UL-Exp, which consists of data lengths different from those used for training and incorporates experimental noise as the noise component, are shown in Fig.~\ref{compare_overall_pointwise_ER_vs_length}(b).
Although UL-Exp differs from UL-Sim in that the noise component is experimental noise, similarly low values of $\mathrm{ER}_{\mathrm{point}}$ comparable to those obtained for UL-Sim are observed across all data lengths.
That is, even under experimental-noise conditions, little to no performance degradation due to variations in data length is observed.

These results indicate that the proposed method exhibits high robustness not only to simulated noise but also to experimentally acquired noise, and that it can maintain stable point-wise transition-detection performance even under conditions where both the data length and the noise environment vary simultaneously.

From these results, it is demonstrated that the proposed method maintains $\mathrm{ER}_{\mathrm{point}}$ at a level comparable to that of TL-Sim even for UL-Sim, which consists of unseen data lengths, without retraining.
In addition, similarly low values of $\mathrm{ER}_{\mathrm{point}}$ are obtained for UL-Exp, where the noise component is experimental noise, confirming that a model trained solely on simulated noise can adapt well to experimental-noise environments.
That is, the proposed method eliminates condition dependence on both data length and noise environment and exhibits high robustness.
However, the evaluations presented so far are primarily conducted under conditions where the noise level is restricted to a fixed range, and the performance variation under changes in the noise intensity itself has not been sufficiently examined.
Therefore, in Sec.~\ref{sec:NL_eval}, the noise robustness of the proposed method is quantitatively evaluated over a wide range of noise levels through a comparison with a conventional threshold-based method.

\section{Noise-Robustness Evaluation through Comparison with a Conventional Threshold-Based Method}
\label{sec:NL_eval}
In this section, the robustness of the U-Net model against variations in the noise level is evaluated through a comparison with a conventional threshold-based method.
The evaluation is conducted using UL-Sim, which consists of simulated noise, and UL-Exp, which consists of experimental noise, and the variation of $\mathrm{ER}_{\mathrm{point}}$ as a function of the noise level is analyzed.

Threshold-based methods are widely used for spin readout signal analysis.
In this study, the threshold is set to half of the signal amplitude.
Because the transition-pulse height is normalized to unity in the datasets used in this study, the threshold is set to 0.5 and transformed using the mean $\mu_{\mathrm{train}}$ and standard deviation $\sigma_{\mathrm{train}}$ of the training data as
\begin{equation}
\theta_{\mathrm{scaled}}
= \frac{0.5 - \mu_{\mathrm{train}}}{\sigma_{\mathrm{train}}} .
\end{equation}
This transformed value is used as the final decision threshold.

\begin{figure}
\begin{center}
  \includegraphics[width=\textwidth]{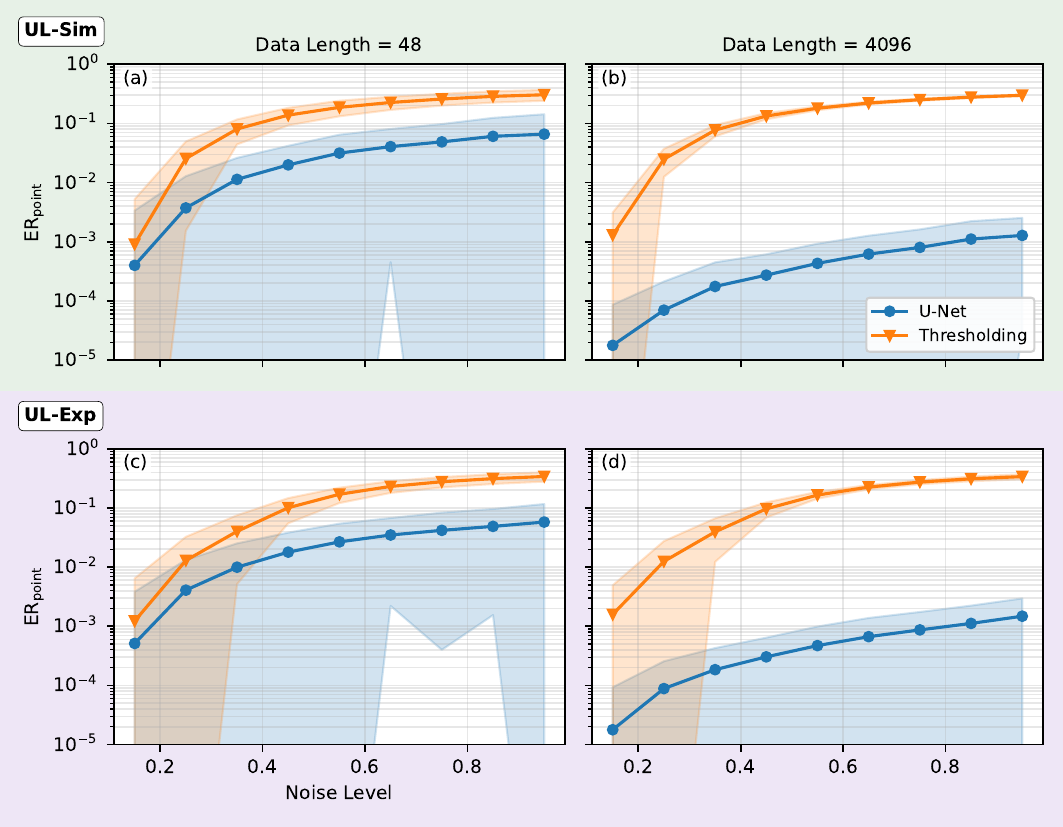}
  \caption{
    Noise-level dependence of the point-wise error rate evaluated on untrained data lengths.
    (a, b) Results for UL-Sim and (c, d) results for UL-Exp.
    The left and right panels correspond to data lengths of 48 and 4096 points, respectively.
    Only samples containing transition events are included.
    Blue and orange curves represent the U-Net and the threshold-based method, respectively.
    Markers indicate mean values and shaded bands represent standard deviations.
    The numbers of samples are approximately 1200 for UL-Sim and 4000 for UL-Exp at each noise level.}
  \label{compare_overall_pointwise_ER_vs_noise_t48_t4096_2x2}
\end{center}
\end{figure}

First, the evaluation is performed on UL-Sim, which is constructed from simulated noise.
Only samples with transition events are used for the evaluation, and two data-length conditions are considered: the shortest length (48 points) and the longest length (4096 points).
The results are shown in Fig.~\ref{compare_overall_pointwise_ER_vs_noise_t48_t4096_2x2}(a) and (b).
For each noise level, the number of samples is approximately 1200, resulting in total numbers of sample points of
$4.9 \times 10^{6}$ (4096 points) and
$5.8 \times 10^{4}$ (48 points), respectively.
As a result, U-Net consistently exhibits lower $\mathrm{ER}_{\mathrm{point}}$ than the threshold-based method, even when the data length and noise level vary.
This behavior is attributed to the fact that U-Net learns local transition shapes as convolutional filters, thereby possessing robustness against variations in both data length and noise intensity.

Next, the results of the same evaluation performed on UL-Exp, which is constructed from experimental noise, are shown in Fig.~\ref{compare_overall_pointwise_ER_vs_noise_t48_t4096_2x2}(c) and (d).
Although UL-Exp differs from UL-Sim in that the noise component is experimental noise, U-Net maintains similarly low values of $\mathrm{ER}_{\mathrm{point}}$ across all noise levels and data lengths, comparable to those obtained for UL-Sim.
That is, no essential difference in noise robustness is observed between the simulation and experimental environments.
Moreover, across the entire noise-level range, U-Net consistently exhibits lower $\mathrm{ER}_{\mathrm{point}}$ than the threshold-based method even for UL-Exp, confirming the superiority of the proposed method under experimental-noise conditions.

In the evaluations presented so far, the classification performance with respect to the temporal locations of transition events has been assessed using the point-wise error rate, $\mathrm{ER}_{\mathrm{point}}$.
In practical spin-state discrimination, however, it is also important to correctly determine whether a transition event is present in an entire trace, that is, the sample-wise classification performance.
Therefore, in Sec.~\ref{sec:sample-wise_eval}, the practical performance of the proposed method is evaluated from the perspective of sample-wise accuracy using UL-Sim and UL-Exp.

\section{Evaluation of Tunnel-Rate–Dependent Sample-Wise Classification Performance through Comparison with a Conventional Threshold-Based Method}
\label{sec:sample-wise_eval}
In this section, the sample-wise classification performance is evaluated from the perspective of the final spin-state discrimination in practical operation.
Here, UL-Sim and UL-Exp are used to assess whether the presence or absence of a transition event can be correctly determined for each sample (i.e., each trace).
As the model outputs prediction probabilities at each point, a trace is classified as a sample with a transition event if at least one point is predicted as an event after point-wise binarization.
As an evaluation metric, the sample-wise accuracy (Acc$_{\mathrm{sample}}$) is employed, which is based on whether the transition event is correctly identified for each sample.
Note that, in the datasets used for these evaluations, the numbers of samples with and without transition events are balanced at a ratio of 1:1 (see Sec.~\ref{sec:eval_data} for details).

The sample-wise accuracy is defined as
\begin{equation}
\mathrm{Acc}_{\mathrm{sample}}
= \frac{TP_{\mathrm{sample}} + TN_{\mathrm{sample}}}
{TP_{\mathrm{sample}} + FP_{\mathrm{sample}} + FN_{\mathrm{sample}} + TN_{\mathrm{sample}}} .
\end{equation}
Here, $TP_{\mathrm{sample}}$ denotes the number of samples with transition events that are correctly classified as having transition events,
and $TN_{\mathrm{sample}}$ denotes the number of samples without transition events that are correctly classified as having no transition events.
The quantity $FP_{\mathrm{sample}}$ represents the number of samples that are incorrectly classified as having transition events despite the absence of true transition events,
while $FN_{\mathrm{sample}}$ represents the number of samples that are incorrectly classified as having no transition events despite the presence of true transition events.
Thus, this metric directly evaluates the correctness of transition-event detection on a per-trace basis.

\begin{figure}
\begin{center}
  \includegraphics[width=\textwidth]{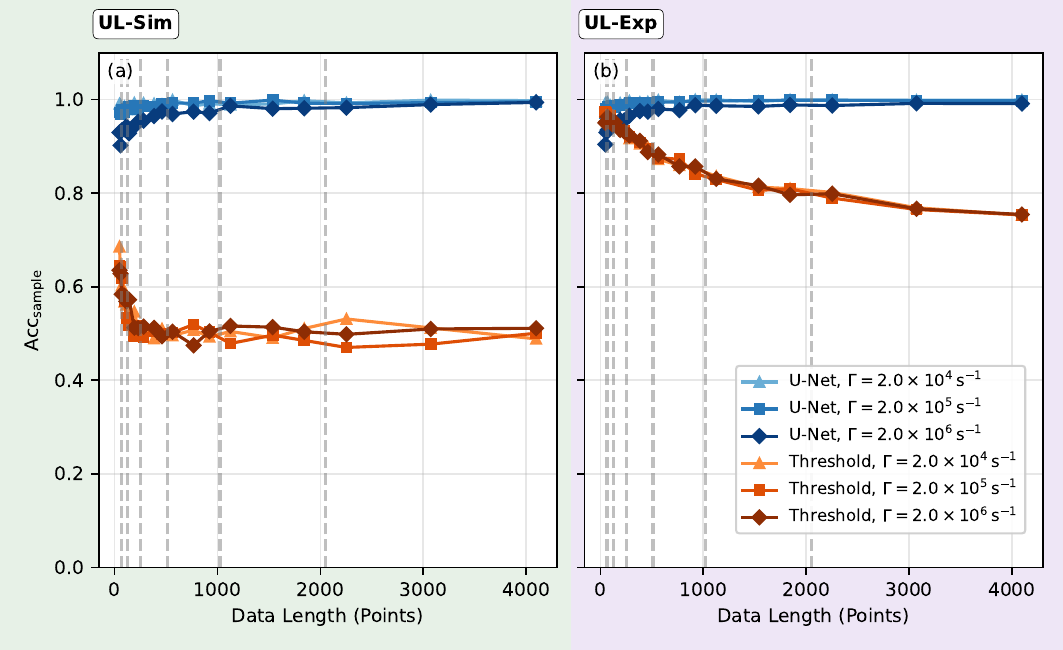}
  \caption{
    Sample-wise accuracy as a function of data length evaluated on
    (a) UL-Sim and (b) UL-Exp.
    The noise level is restricted to $0.2 \leq \text{noise level} < 0.3$.
    Blue and orange curves correspond to the U-Net and the threshold-based method, respectively,
    with color intensity indicating different tunneling rates $\Gamma$.
    The gray dashed lines denote the data lengths used during training.}
  \label{compare_samplewise_Acc_perTR_vs_length}
\end{center}
\end{figure}

First, the data-length dependence of the sample-wise classification performance is evaluated using UL-Sim, which consists of simulated noise, and the results are shown in Fig.~\ref{compare_samplewise_Acc_perTR_vs_length}(a).
As the tunneling rate $\Gamma$ varies, the time scale of electron tunneling changes, resulting in variations in the temporal width of transition events.
In particular, at faster tunneling rates, electron tunneling occurs more rapidly, and the resulting transition events become extremely narrow and appear only momentarily in time, making them inherently difficult to identify.
Nevertheless, for all data-length and tunneling-rate conditions, U-Net consistently achieves higher $\mathrm{Acc}_{\mathrm{sample}}$ than the threshold-based method.
This indicates that U-Net maintains stable sample-wise classification performance even when both the data length and tunneling rate vary.
In contrast, for the threshold-based method, $\mathrm{Acc}_{\mathrm{sample}}$ exhibits a tendency to decrease toward approximately 0.5 as the data length increases.
This behavior indicates that, as the data length becomes longer, the probability that noise exceeds the threshold increases, causing the discrimination performance for the presence or absence of transition events to degrade to a level comparable to random guessing.
These results suggest that the threshold-based method is strongly affected by stochastic false detections associated with increasing data length.

Next, the data-length dependence of the sample-wise classification performance is evaluated in the same manner for UL-Exp, which is constructed from experimental noise, and the results are shown in Fig.~\ref{compare_samplewise_Acc_perTR_vs_length}(b).
In the UL-Exp results (Fig.~\ref{compare_samplewise_Acc_perTR_vs_length}(b)), the sample-wise accuracy of the threshold-based method is overall slightly higher than that observed for UL-Sim (Fig.~\ref{compare_samplewise_Acc_perTR_vs_length}(a)).
This difference is attributed to the fact that the amplitude distribution of the experimental noise used in UL-Exp is asymmetric and biased toward values below the mean (see Sec.~\ref{sec:UL-Exp_preparation} for details), making it statistically less likely for noise fluctuations to exceed a fixed threshold.
As a result, this noise characteristic leads to conditions that are inadvertently favorable for the threshold-based method.
If the noise distribution were instead biased toward values above the mean, the threshold-based method would experience an increased number of false detections, potentially leading to a substantial degradation in sample-wise classification performance.
This highlights the strong sensitivity of threshold-based analysis to biases in the noise distribution.
By contrast, U-Net maintains a sample-wise accuracy comparable to that obtained for UL-Sim even under experimental-noise conditions whose statistical properties differ from those of simulated noise.
By basing its classification on the local temporal trace structures of transition events, U-Net achieves stable performance that is robust against variations in the noise distribution.

From these results, it is demonstrated that although the threshold-based method may exhibit temporarily high performance under accidentally favorable noise-distribution conditions, its performance is inherently unstable.
In terms of reliability in real experimental environments, U-Net is therefore shown to be clearly superior.

\section{Conclusion}
This paper has presented a U-Net framework for spin readout signal analysis that addresses fundamental limitations of conventional threshold-based and neural-network-based approaches.
By reformulating spin readout as a point-wise transition-event detection problem in one-dimensional time-series data, the proposed method simultaneously mitigates condition dependence on training data, fixed-length input constraints, and the lack of temporal resolution inherent to trace-level classification models.

A central contribution of this study is the demonstration that a fully convolutional U-Net architecture enables direct and accurate localization of transition events while naturally supporting variable-length input traces.
Training on a large-scale simulated dataset spanning a wide range of noise levels, tunneling rates, and data lengths allows the model to acquire robust and condition-independent representations without requiring retraining or manual parameter tuning.
As a result, stable performance is maintained not only for data lengths used during training but also for previously unseen lengths and experimental-noise conditions.

Comprehensive evaluations from both point-wise and sample-wise perspectives confirm the practical effectiveness of the proposed approach.
Point-wise evaluations show that consistently low error rates are achieved across variations in data length and noise level, demonstrating that the proposed method can produce temporally resolved outputs that explicitly indicate the locations of transition events within a trace with high accuracy, which cannot be obtained using conventional trace-level classification.

Sample-wise evaluations further reveal that the proposed method achieves stable and superior spin-state discrimination performance compared to a conventional threshold-based method, even under fast tunneling-rate and experimental-noise conditions.

Taken together, these results establish that the proposed U-Net method provides a unified and robust solution for spin readout signal analysis across diverse experimental conditions.
Notably, no fundamental difference in noise robustness is observed between UL-Sim and UL-Exp, despite their different noise distributions.
Since noise distributions in practical experiments often vary depending on experimental conditions, this result indicates that the proposed method is expected to adapt well to such variations in real spin readout measurements.
This robustness is likely attributable to the ability of the U-Net to capture characteristic transition-event patterns across variations in data length and noise characteristics.
By combining temporally resolved outputs, compatibility with variable-length traces, and strong generalization to experimental noise, the proposed framework enables practical automation of spin readout analysis in real measurement environments.
This work represents an important step toward improving the accuracy and reliability of measurements in semiconductor quantum-dot–based quantum computing systems, and contributes to the broader goal of automating qubit operations through spin readout signal analysis.

\bibliography{references}

\section*{Acknowledgements}
Part of this work is supported by 
Grants-in-Aid for Scientific Research (23KJ0200, 23K26482, 23H04490, 23H05455, 25H01504),
JST FOREST (JPMJFR246L),
JST CREST (JPMJCR23A2),
Mitsubishi Foundation Research Grant,
Yasumi Science and Technology Foundation Research Grant, 
The Kazuchika Okura Memorial Foundation Research Grant,
HABATAKU Young Researchers Support Program,
FRiD Tohoku University.
Y. M. acknowledges WISE Program of AIE for financial support.

\end{document}